\documentclass[aps,pra,nobalancelastpage,superscriptaddress, twocolumn,10pt]{revtex4-2}
\usepackage{color,ifthen,amsthm,amsmath,amsxtra,amsfonts,dsfont,pgfplots,graphicx,bm,tikz,scalerel,wasysym,bbm,graphicx,amsthm,braket,physics,empheq,CircuitTikz}
\usepackage[colorlinks=true,linkcolor=blue, citecolor=blue, urlcolor=blue, bookmarks]{hyperref}

\newcommand{\be}{\begin{equation}}
\newcommand{\ee}{\end{equation}}
\newcommand{\bea}{\begin{eqnarray}}
\newcommand{\eea}{\end{eqnarray}}
\newcommand{\bse}{\begin{subequations}}
\newcommand{\ese}{\end{subequations}}

\theoremstyle{plain}

\theoremstyle{plain}

\theoremstyle{plain}

\newcommand{\brakapp}{^{(\prime)}}
\newcommand{\bU}{\mathbb{U}}

\newcommand{\avg}[1]{\left\langle #1\right\rangle}
\newcommand{\floor}[1]{\left\lfloor#1\right\rfloor}
\newcommand{\ceil}[1]{\left\lceil#1\right\rceil}

\newcommand{\aU}{\mathcal{W}}
\newcommand{\spr}[2]{\bra{#1}#2\rangle}

\renewcommand{\O}{\mathcal{O}}
\renewcommand{\norm}[1]{\left\|#1\right\|}

\definecolor{favg}{rgb}{0.4,0.4,0.86}
\definecolor{flines}{rgb}{0.3,0.3,0.3}

\begin{document}
\title{Initial State Memory in Finite Random Brickwork Circuits}

\author{Jakob Bannister}
\affiliation{School of Physics and Astronomy, University of Birmingham, Edgbaston, Birmingham, B15 2TT, UK}

\author{Katja Klobas}
\affiliation{School of Physics and Astronomy, University of Birmingham, Edgbaston, Birmingham, B15 2TT, UK}

\author{Colin Rylands}
\affiliation{Centre for Fluid and Complex Systems, Coventry University, Coventry, CV1 2TT, UK}

\author{Bruno Bertini}
\affiliation{School of Physics and Astronomy, University of Birmingham, Edgbaston, Birmingham, B15 2TT, UK}

\begin{abstract}
We ask under what conditions a finite brickwork circuit of random gates retains local information about the initial state. To answer this question we measure the averaged Frobenius distance between the reduced states obtained by evolving two arbitrary initial states and tracing out a portion of the system. By characterising this distance exactly at all times we find that the information is retained if the environment --- the subsystem traced out --- is smaller than half of the system and washed away otherwise. We also find that, while the dynamics of the Frobenius distance depends on the specific initial states chosen, this dependence becomes increasingly weak for large scales and eventually the Frobenius distance attains a universal form as a function of time. Finally, we show that by introducing weak enough boundary dissipation, one can observe a phase transition between a memory preserving phase and one where the information is completely lost.

\end{abstract}

\maketitle

\section{Introduction}
Closed quantum systems evolve via unitary transformations, perfectly retaining all information encoded in their initial state. However, local information of the initial state is generically dispersed (or `scrambled') throughout the full system so that from the point of view of a small subsystem the initial state is forgotten~\cite{hayden2007black, sekino2008fast, swingle_unscrambling_2018}. In the context of many-body physics, this local memory loss leads to the emergence of universality in late time quantum dynamics~\cite{polkovnikov2011colloquium, calabrese2016introduction, eisert2015quantum, gogolin2016equilibration, bastianello2022introduction}. Fine grained details of an initial state are washed away in favour of coarse grained information set by conservation laws, leading to the onset of hydrodynamics and the emergence of statistical mechanics~\cite{castroalvaredo2016emergent,bertini2016transport,deutsch1991quantum,srednicki1994chaos, calabrese2016introduction, eisert2015quantum, gogolin2016equilibration, bastianello2022introduction}. A new symmetry-based perspective on this phenomenon has recently been  widely investigated wherein one studies dynamical symmetry restoration from symmetry broken initial states at the level of a subsystem~\cite{ares2022entanglement, ares2025quantum}. In this setting, local information about the symmetry breaking is transported throughout the system via conserved currents~\cite{rylands2024microscopic,murciano2024entanglement}. Memory of the symmetry breaking is thus forgotten within the subsystem, allowing the symmetry to be restored apart from some exceptional cases~\cite{ares2023lack}. Interestingly, the symmetry need not be an exact symmetry of the dynamics but can be emergent. One example of this is the restoration of space-time or internal symmetry under local  random unitary dynamics, which acquires these symmetries only upon averaging over realisations~\cite{klobas2025translation,ares2024entanglement,russotto2025non}. 

Within quantum information theory, the loss of initial state memory is connected with the preparation of random states which are highly desirable for certain quantum computation algorithms~\cite{dankert2009exact}. Starting from some simple initial state, one would like to prepare, as efficiently as possible, a state which approximates a random state to a certain accuracy. Random unitary circuits have been a key player in this regard as they are known to form $k$-designs in polynomial time, i.e. the can match the $k^{\rm th}$ moment of a Haar random state~\cite{harrow2009random, brandao2016local, brandao2016efficient, schuster2025strong}. From the symmetry perspective, Haar random states are, on average, completely symmetric. Thus the loss of initial state memory and the preparation of random states are connected with the restoration of the emergent symmetries in random unitary circuits. 

In this paper we investigate whether random unitary dynamics can erase the memory of an initial state at the subsystem level, and if so, on what time scales. Specifically, we follow the setting of Ref.~\cite{klobas2025translation}: we imagine two different (pure) initial states, which are both undergoing the \emph{same} time-evolution. At some later time $t$ we measure the distance between the two states reduced to a \emph{local} subsystem. The initial states might be related via symmetry, in which case this protocol is probing the ability of the dynamics to locally restore the symmetry. More generally, however, the initial states can be unrelated, in which case we probe how memory of the initial state is retained over time.

For pure states we find that if the size of the subsystem is less than half of the total system size the two reduced states become exponentially close to each other on time scales of the order of the subsystem size. For subsystems larger than half of the total system size, we show that the two time-evolved states do not approach each other, which means that the memory of the initial state is retained for \emph{all} times. Moreover, for appropriately chosen initial states the final states can become maximally distant even if they start close together. For mixed states the transition between memory loss and retention as a function of subsystem size can be shifted away from half-system-size to larger values.

We also investigate the effect of non-unitary dynamics adding dissipation to one of its boundaries. The effect of this will be to erase all memory of the initial state, and therefore drive symmetry restoration, regardless of subsystem size. We show, however, that for noise which decays algebraically in time, memory of the initial state can be retained up to large times. We find an expression similar to that of mixed states and determine the critical value of dissipation at which the transition occurs. 

The remainder of the article is structured as follows. In Sec.~\ref{sec:setting} we introduce the setting and the quantities that we consider throughout the paper. In Sec.~\ref{sec:exactannealedaverage} we find the exact expression for the Frobenius norm of the distance between the two time-evolved states, averaged over disordered random circuits. We analyse its behaviour in different regimes and limits, considering different classes of initial states. In Sec.~\ref{sec:OpenDynamics} we complement this picture by adding dissipation to the boundary degrees of freedom, which aids the loss of the initial-state memory even when subsystem size is large compared to the rest. However, by appropriately scaling the strength of the dissipation with the system size and time, one we find a nontrivial transition between memory loss and retention. Finally, in Sec.~\ref{sec:conclusions} we conclude with some closing remarks and open questions. A number of technical details are delegated to appendices.

\section{Setting}\label{sec:setting}
We consider a brickwork quantum circuit of $2L$ qudits with local Hilbert space dimension $q$.  The system is initialised in a non-equilibrium state $\ket{\Psi}$ and evolves with time evolution operator $\bU$ as $\ket{\Psi(t)}=\bU^t \ket{\Psi}$ with $t\in\mathbb N$. We  ask whether the state of a local subsystem $A$ eventually forgets all the information about the initial state and, if it does, what is the timescale for this process to occur. To answer this question we compare the dynamics of two different states, $\ket{\Psi}$ and $\ket{\Psi'}$, and measure the distance between the reduced density matrices $\rho_A(t) = \tr_{\bar{A}}(\ket{\Psi(t)}\!\bra{\Psi(t)})$ and $\rho'_A(t) = \tr_{\bar{A}}(\ket{\Psi'(t)}\!\bra{\Psi'(t)})$ by computing the normalised Frobenius norm of their difference 
\begin{equation}
\label{eq:frobeniusdistance}
    \Delta_2(t)=\frac{\norm{\rho_A(t)-{\rho}'_A(t)}_2}{\norm{\rho_A(t)}_2+\norm{\rho'_A(t)}_2},
\end{equation}
where $\norm{A}_2 = \sqrt{\tr[AA^\dagger]}$. Note that $\Delta_2(t)$ is symmetric, non-negative, and, importantly, $\Delta_2(t)=0$ if and only if $\rho_A(t)={\rho}'_A(t)$. Therefore, if the information about the initial state contained within the subsystem is eventually completely forgotten, the two reduced density matrices will at some time $t^{\prime}$ be indistinguishable, and $\Delta_2(t^{\prime})=0$. If, however, dependence on the initial state is retained, then, necessarily $\Delta_2(t)>0$.

We consider a brickwork with the time evolution operator written as 
\begin{equation}
\label{eq:timeevolutionop}
    \bU = \smashoperator{\bigotimes_{j=1}^{L}} U_{j,j+\frac{1}{2}} \smashoperator{\bigotimes_{j=2}^L}U_{j-\frac{1}{2},j}, 
\end{equation}
where the local gates $U_{j,j+\frac{1}{2}}$ are unitary matrices acting only on the sites $j$ and $j+{1}/{2}$. We label qudit positions by half integers and, although we did not make it explicit in our notation, we focus on a case where the local gates depend on both position and time. Finally for concreteness, in Eq.~\eqref{eq:timeevolutionop} we have taken open boundary conditions, but our results can be directly extended to periodic boundary conditions. 

We wish to study the effects of the locality of the interactions and not of the special properties of the local gates, therefore in this paper we consider each local gate to be a random unitary matrix following independent Haar distributions. Our interest will then be in averaged quantities, which we denote with angled brackets $\langle\cdot\rangle$, and specifically on the average of Eq.~\eqref{eq:frobeniusdistance}. We note that the key features of $\Delta_2(t)$ (symmetry, non-negativity, and the fact that $\Delta_2(t)=0$ if an only if $\rho_A(t) ={\rho}'_A(t)$) are shared by its annealed version 
\begin{equation} \label{eq:annealedaverage}
  \avg{\Delta_2(t)}_a=\sqrt{\frac{\expval{\tr((\rho_A(t)-{\rho}'_A(t))^2)}}{\expval{\tr(\rho_A(t)^2)}+\expval{\tr({\rho}'_A(t)^2)}}},
\end{equation}
where we have taken the Haar average inside the square root. We will focus on this simpler quantity from now on. Expanding the product, one may rewrite the square of this quantity as  
\begin{equation}
\label{eq:annealedaveragesquared}
    \avg{\Delta_2(t)}_a^2 = 1-\frac{2\avg{\tr[\rho_A(t)\rho'_A(t)]}}{\avg{\tr[\rho_A(t)^2]}+\avg{\tr[\rho'_A(t)^2]}}.
\end{equation}
Noting that for any pair of density matrices $\rho$ and $\rho'$ we have 
\be
0 \leq 2 \tr[\rho \rho'] \leq \tr[\rho^2]+\tr[\rho'^2], 
\ee
where the first inequality follows from the positivity of the reduced density matrices and the second by their Hermiticity, we find that  
\be
\label{eq:boundsDelta}
0 \leq \avg{\Delta_2(t)}_a^2 \leq 1\,. 
\ee
Moreover, from Eq.~\eqref{eq:annealedaveragesquared} we see that to evaluate the Frobenius distance it is sufficient to calculate $\expval{\smash{\tr[\smash{\rho_A(t)\rho\brakapp_A(t)}]}}$, where $(\prime)$ means that we can chose to have the prime symbol or not. This object is easily represented graphically as follows 
\begin{equation} \label{eq:traceproduct}
  \tr[\rho_A(t)\rho_A\brakapp(t)] = 
  \begin{tikzpicture}[baseline={(0,-1.85)},scale=0.5]
    \foreach \x in {-5,-3,...,3}{\prop{\x}{0}{favg}{}}
    \foreach \x in {-4,-2,...,2}{\prop{\x}{-1}{favg}{}}
    \foreach \x in {-5,-3,...,3}{\prop{\x}{-2}{favg}{}}
    \foreach \x in {-4,-2,...,2}{\prop{\x}{-3}{favg}{}}
    \foreach \x in {-5,-3,...,3}{\prop{\x}{-4}{favg}{}}
    \foreach \x in {-4,-2,...,2}{\prop{\x}{-5}{favg}{}}
    \foreach \x in {-5.5,...,-3.5}{\circle{\x}{0.5}}
    \foreach \x in {-2.5,...,3.5}{\square{\x}{0.5}}
    \draw[color = flines, line width = 0.2mm](-5.5,-.5) -- (-5.5,-1.5);
    \draw[color = flines, line width = 0.2mm](3.5,-2.5) -- (3.5,-3.5);
    \draw[color = flines, line width = 0.2mm](-5.5,-2.5) -- (-5.5,-3.5);
    \draw[color = flines, line width = 0.2mm] (3.5,-.5) -- (3.5,-1.5);
    \draw[color = flines, line width = 0.2mm] (-5.5,-4.5) -- (-5.5,-5.5);
    \draw[color = flines, line width = 0.2mm] (3.5,-4.5) -- (3.5,-5.5);
    \draw[color = flines, line width = 0.2mm, fill=colSt] (-5.55,-6.6) rectangle (3.65,-5.5) node[pos=.5] {$\Psi_4\brakapp$};
    \draw[semithick,decorate,decoration={brace}] (-2.65,.76) -- (3.65,.76) node[midway,xshift=0pt,yshift=7.5pt] {$A$};
  \end{tikzpicture}.
\end{equation}
Here we have depicted the averaged, folded, replicated, local gate, $\aU_2$,  as 
\begin{equation} \label{eq:foldedgate}
  \aU_2=\avg{U\otimes U^*\otimes U\otimes U^*}=
  \begin{tikzpicture}[baseline={(0,-0.1)},scale=.6]
    \prop{0}{0}{favg}{}
  \end{tikzpicture}. 
\end{equation}
We take the subsystem $A$ to consist of the set of contiguous qudits starting from the right boundary. This choice imposes a particular boundary condition at the top of our diagram which is depicted with the help  of the following special states in the replicated space
\be
\label{eq:squarecirclestates}
    \ket{\circleSA} = \sum_{r,s=1}^q\ket{r,r,s,s}, \quad
    \ket{\squareSA} = \sum_{r,s=1}^q \ket{r,s,s,r}~. 
\ee
These represent traces which can be performed within each replica ($\circleSA$) or after coupling different replicas first ($\squareSA$).   Lastly, the boundary condition at the bottom comes from the initial states in the replicated space which we denote by 
\begin{equation}
  \ket*{\Psi_4\brakapp} = 
  \ket*{\Psi}\otimes\ket*{\Psi}^*\otimes\ket*{\Psi\brakapp}\otimes\ket*{\Psi\brakapp}^*.
\end{equation}
We consider pure states but the formalism is directly generalised to mixed states.

It is straightforward to show that, because of the unitarity of the local gates, the special states in Eq.~\eqref{eq:squarecirclestates} and the folded gate in Eq.~\eqref{eq:foldedgate} satisfy the following relations
\begin{equation} \label{linking_state_rules_1}
  \aU_2\ket{\circleSA\,\circleSA} = 
  \mkern-8mu
  \begin{tikzpicture}[baseline={(0,-0.1)},scale=.475]
    \prop{0}{0}{favg}{}\circle{0.5}{0.5}\circle{-0.5}{0.5}
  \end{tikzpicture}
  \mkern-8mu
  =\ket{\circleSA\,\circleSA},\quad
  \aU_2\ket{\squareSA\,\squareSA} = 
  \mkern-8mu
  \begin{tikzpicture} [baseline={(0,-0.1)},scale=.475]
    \prop{0}{0}{favg}{}\square{0.5}{0.5}\square{-0.5}{0.5}
  \end{tikzpicture}
  \mkern-8mu
  =\ket{\squareSA\,\squareSA}.
\end{equation}
In addition, as a result of the averaging with Haar measure, we have 
\begin{equation}\label{linking_state_rules_2}
  \mkern-16mu
  \aU_2\ket{\circleSA\,\squareSA} =\mkern-8mu
  \begin{tikzpicture}[baseline={(0,-0.1)},scale=.475]
    \prop{0}{0}{favg}{}\circle{-0.5}{0.5}\square{0.5}{0.5}
  \end{tikzpicture} \mkern-8mu
  %= \alpha(\ket{\circleSA\,\circleSA}+\ket{\squareSA\,\squareSA}),\\
  =\aU_2\ket{\squareSA\,\circleSA} =
  \mkern-8mu
  \begin{tikzpicture}[baseline={(0,-0.1)},scale=.475]
    \prop{0}{0}{favg}{}\circle{0.5}{0.5}\square{-0.5}{0.5}
  \end{tikzpicture}\mkern-8mu= \alpha(\ket{\circleSA\,\circleSA}+\ket{\squareSA\,\squareSA}),
  \mkern-16mu
\end{equation}
where we introduced
\begin{equation}
  \alpha=\frac{q}{q^2+1}. 
\end{equation}
In the following section we will show how Eq.~\eqref{eq:traceproduct} (and hence Eq.~\eqref{eq:annealedaverage}) can be efficiently computed exactly for the setting at hand. 

\section{Annealed average under unitary time-evolution}
\label{sec:exactannealedaverage}

Let us now consider an exact calculation of the annealed average of $\Delta_2(t)$ for (so far) arbitrary initial states. For system sizes that are much larger than time $t$ and uncorrelated initial states we recover the results of Ref.~\cite{klobas2025translation}, stating that --- up to exponentially small corrections in the subsystem size --- all the initial-state dependence disappears at times proportional to the subsystem size.

We expect this behaviour to change when the ratio between the subsystem and the system sizes becomes finite. To illustrate this, let us for a moment consider the infinite-time value of $\avg{\Delta_2(t)}_a$. For $L\to\infty$ we expect $\rho_{A}(t)$ to converge to the infinite temperature state (regardless of the initial state) for all realisations of our random circuit except zero measure sets. Denoting the subsystem size by $x$ this gives the following expectations
\begin{equation}
  %\begin{aligned}
    \lim_{t\to\infty}\lim_{L\to\infty}
    \expval{\tr\smash{[\smash{\rho_A(t)\rho_A^{\prime}(t)}]}}=
    \expval{\tr\smash{[\smash{\rho^{(\prime)}_A(t)^2}]}}=
    \frac{1}{q^{x}}, 
    %&= \frac{1}{q^{x}},\\
    %\lim_{t\to\infty}\lim_{L\to\infty}\avg{\tr[\rho_A(t)^2]} &= \frac{1}{q^{x}},\\
    %\lim_{t\to\infty}\lim_{L\to\infty}\avg{\tr[\rho'_A(t)^2]} &= \frac{1}{q^{x}},
  %\end{aligned}
\end{equation}
and hence
\begin{equation}
  \lim_{t\to\infty}\lim_{L\to\infty}\avg{\Delta_2(t)}_a^2=0.
\end{equation}
For finite systems, however, there is no reason for this to happen. In fact, Page's formula~\cite{page1993average} implies that even when the state of the system is fully random its reduced state on a subsystem $A$ is maximally mixed only if $A$ is smaller than half of the system. In essence this is the case because for $x\equiv |A|<L$, $A$ is maximally entangled with a subsystem of $\bar A = \{1/2,\ldots, L\}\setminus A$, while for $x>L$ $\bar A$ is maximally entangled with a subsystem of $A$~\cite{page1993average, hayden2007black, xu2024scrambling}.

This suggests that, for ${L< x \leq 2 L}$, the infinite-time value of $\avg{\Delta_2(t)}_a$ is non-zero (an example of this, albeit for other symmetries, was also observed in Refs.~\cite{ares2024entanglement,russotto2025non,ares2025entanglement}). As a consequence, also the finite-time dynamics should behave qualitatively differently when $x\geq L$ than it does for $x\leq L$, which is what we consider in this section. To investigate these questions we start by finding an exact expression for the annealed average in Sec.~\ref{sec:ExactExpression}, which we then compare with the above $t\to\infty$ expectation in Sec.~\ref{sec:InfiniteTime}, and finally in Sec.~\ref{sec:IntermediateTime} we characterise its finite-time behaviour for a number of representative initial states.

\subsection{Exact expression for the annealed average}
\label{sec:ExactExpression}

Our strategy to exactly contract  the diagram in Eq.~\eqref{eq:traceproduct} is to use Eq.~\eqref{linking_state_rules_1} and Eq.~\eqref{linking_state_rules_2} to propagate  the top layer  of the circuit diagram downwards. Specifically,
% denoting the size of $A$ by $x$--- which we take to be odd --- 
taking $x$ to be odd,
we write the state corresponding to a horizontal cut $m$ steps away from the top as 
\begin{equation} \label{eq:Tdef}
  \bra{T_m(x)}:= \bra{\circleSA^{2L-x}\,\squareSA^x}\mathcal{U}_m. 
\end{equation}
Here $\mathcal{U}_m$ denotes $m$ rows of time-evolution, depicted diagrammatically below for $m=6$
\begin{equation}
  \mathcal{U}_m = 
  \begin{tikzpicture}[baseline={(0,-1.4)},scale=0.5]
    \foreach \x in {-9,-7,...,-1}{\prop{\x}{0}{favg}{}}
    \foreach \x in {-8,-6,...,-2}{\prop{\x}{-1}{favg}{}}
    \foreach \x in {-9,-7,...,-1}{\prop{\x}{-2}{favg}{}}
    \foreach \x in {-8,-6,...,-2}{\prop{\x}{-3}{favg}{}}
    \foreach \x in {-9,-7,...,-1}{\prop{\x}{-4}{favg}{}}
    \foreach \x in {-8,-6,...,-2}{\prop{\x}{-5}{favg}{}}

    \draw[color = flines, line width = 0.2mm](-9.5,-.5) -- (-9.5,-1.5);
    \draw[color = flines, line width = 0.2mm](-.5,-2.5) -- (-.5,-3.5);
    \draw[color = flines, line width = 0.2mm](-9.5,-2.5) -- (-9.5,-3.5);
    \draw[color = flines, line width = 0.2mm] (-.5,-.5) -- (-.5,-1.5);
    \draw[color = flines, line width = 0.2mm] (-9.5,-4.5) -- (-9.5,-5.5);
    \draw[color = flines, line width = 0.2mm] (-.5,-4.5) -- (-.5,-5.5);
    \draw [decorate,decoration={brace,amplitude=10pt,raise=4pt},yshift=0pt]
    (0,0.5) -- (0,-5.5) node [black,midway,xshift=0.8cm] {$m$};
  \end{tikzpicture}.
\end{equation}
Note that this corresponds to $t=m/2$ time steps for $m$ even, and for odd $m$ the bottom-most row changes (i.e.\ the parity of the top row is fixed). Therefore, the diagram in Eq.~\eqref{eq:traceproduct} can be written as 
\begin{equation} \label{eq:productT}
  \expval*{\tr[\smash{\rho_A(t)\rho_A\brakapp(t)}]} = \braket*{T_{2t}(x)}{\Psi_4\brakapp}. 
\end{equation}
By virtue of Eq.~\eqref{linking_state_rules_2} the state in Eq.~\eqref{eq:Tdef} fulfils 
\be
\bra{T_m(x)} = \alpha \left[\bra{T_{m-1}(x-1)}+\bra{T_{m-1}(x+1)}\right],
\label{eq:Tstaterel}
\ee
unless $x=0$ or $2L$. In either of these special cases, the top state is  composed only of $\circleSA$ or $\squareSA$-states and the circuit acts on it as the identity. Otherwise, Eq.~\eqref{eq:Tstaterel} means that the domain wall between $\circleSA$ and $\squareSA$ states prepared at $m=0$ performs a random walk that terminates when it reaches the boundaries of our system (this is consistent with the statistical mechanics mapping of Refs.~\cite{zhou2019emergent, nahum2018operator}).

Applying this relation $m$ times, we decompose $\bra{T_m(x)}$, as follows
\be
\label{eq:Tstateexpanded}
\begin{aligned}
    \bra{T_m(x)} =& A_m(x)\bra{T_0(0)}+B_m(x)\bra{T_0(2L)} \\
    &+\alpha^m\sum_{k=k_{\min}}^{k_{\max}}C_{k,m}(x)\bra{T_0(x-m+2k)},
\end{aligned}
\ee
where $k$ represents the number of rightward steps our domain wall takes through the system and we set 
\begin{equation}
  \begin{aligned}
    k_{\min}&=\max\left(0,\ceil{\frac{m-x}{2}}\right),\\
    k_{\max}&=\min\left(m,\floor{\frac{m-x+2L}{2}}\right).
  \end{aligned}
\end{equation}
In writing Eq.~\eqref{eq:Tstateexpanded} we conveniently separated the stationary states $\bra{T_0(0)}$ and $\bra{T_0(2L)}$ from the rest. The coefficients of the non-stationary parts are easily seen from Eq.~\eqref{eq:Tstaterel} to fulfil the recursive relation
\be
\label{eq:binomial}
C_{k,m+1}(x)=C_{k,m}(x)+C_{k-1,m}(x), 
\ee
with the boundary conditions
\be
\label{eq:boundary}
C_{k_{\min}-1,m}(x)=C_{k_{\max}+1,m}(x)=0. 
 \ee
% \be
% \label{eq:boundary}
% C_{\ceil{\frac{m-x}{2}}-1,m}(x)=C_{\floor{\frac{m-x+2L}{2}}+1,m}(x)=0. 
% \ee
Eq.~\eqref{eq:binomial} is just a binomial relation, and our boundary conditions can be satisfied by adding simple corrections involving appropriate sums of binomial coefficients. Specifically, one can show that Eqs.~\eqref{eq:binomial} and~\eqref{eq:boundary} are solved by
\begin{equation}
  C_{k,m}(x) = {m\choose k}-{m\choose k+x}-r_{k,m}(x),
\end{equation}
where
\begin{equation} \label{eq:rdef}
  \begin{aligned}
    r_{k,m}(x) = \sum_{n=1}^\infty&\left[\binom{m}{k+x+2nL}+{m \choose k+x-2nL}\right.\\
    &\left.-\binom{m}{k+2nL}-\binom{m}{k-2nL}\right].
  \end{aligned}
\end{equation}
To find $A_m(x)$ we note that the latter is given by the sum of all the coefficients $C_{k,m}(x)$ of the $\bra{T_0(x)}$ terms touching the boundary $x=0$ for $m' = 0, \ldots, m$. This gives 
\begin{equation}\label{eq:ExpressionForAm}
 A_m(x) = \alpha^x\sum_{k=0}^{k_{\min}}\alpha^{2k}C_{k,2k+x-1}(x).
\end{equation}
Analogously we find 
\begin{equation}\label{eq:A_B_relation}
B_m(x)= A_m(2L-x).
\end{equation}
Note that, in contrast with $C_{k,m}(x)$, these coefficients are not exponentially suppressed in time, and thus dominate the large-time regime. 

Plugging Eq.~\eqref{eq:Tstateexpanded} into Eq.~\eqref{eq:productT} gives a closed form expression for $\expval{\tr\smash{[\smash{\rho_A(t)\rho_A\brakapp(t)}]}}$ and hence, via Eq.~\eqref{eq:annealedaveragesquared}, for $\avg{\Delta_2(t)}_a$. The resulting expression we find for $\avg{\Delta_2(t)}_a$, however, is not particularly transparent as it involves a ratio of double sums of binomial coefficients. In the following, we proceed to simplify it in relevant special cases.

\subsection{Infinite time regime}\label{sec:InfiniteTime}
As discussed at the beginning of Sec.~\ref{sec:exactannealedaverage}, we expect that the stationary value of 
\be
\lim_{t\to\infty}\avg{\Delta_2(t)}_a^2, 
\ee
approaches $0$ for subsystem sizes $x$ that are small compared to $2L$ (more precisely, $x<L$), and might approach a non-zero value for $L<x<2L$. Let us test this expectation using our exact result in Eq.~\eqref{eq:Tstateexpanded}.

First we note that, since $C_{k,m}(x)$ are exponentially suppressed in the number of time steps $m$, they do not contribute to the infinite-time limit. Thus, we only need to calculate $A_\infty(x)$ and $B_\infty(x)$.

As we show in App.~\ref{sec:AppendixAinfty}, the $m\to\infty$ limit of Eq.~\eqref{eq:ExpressionForAm} can be simplified to
\begin{equation}
  \label{eq:Ainf}
  A_\infty(x) = q^{-x} -\frac{2\sinh{(x\ln q)}}{q^{4L}-1}\simeq q^{-x}. 
\end{equation}
where $\simeq$ denotes the leading contribution for large $L$. The other coefficient, $B_{\infty}(x)$, is obtained from $A_{\infty}(x)$ through Eq.~\eqref{eq:A_B_relation} and reads
\begin{equation} 
\label{eq:Binf}
\!\!\!\!B_\infty(x)=q^{-2L+x}\!-\!\frac{2\sinh{((2L-x)\ln q)}}{q^{4L}-1}\simeq q^{-2L+x}. 
\end{equation}
Note also that --- as shown in App.~\ref{sec:alltoall} --- the result coincides with that obtained by replacing our time evolution operator $\bU$ with an all-to-all random unitary matrix. This is consistent with the fact that, for infinite times, brickwork quantum circuits become strong $k$-designs~\cite{schuster2025strong,harrow2009random,brandao2016local,brandao2016efficient}. 

Plugging Eqs.~\eqref{eq:Ainf} and \eqref{eq:Binf} back into Eq.~\eqref{eq:Tstateexpanded}, and using Eqs.~\eqref{eq:productT} and~\eqref{eq:annealedaveragesquared} we finally obtain 
\begin{equation}\label{eq:long_time_full}
    \lim_{t\to\infty}\avg{\Delta_2(t)}_a^2 \simeq 1-\frac{q^{-x}+|\spr{\Psi}{\Psi'}|^2 q^{-2L+x}}{q^{-x}+q^{-2L+x}}, 
\end{equation}
where we neglected terms that are sub-leading for large $L$. Taking the limit $x,L\to\infty$ with fixed ratio $r=x/(2L)$ we get two distinct limits depending on whether $r$ is smaller or larger than $1/2$. Namely we find 
\begin{equation}\label{eq:long_time_behaviour}
   \lim_{\substack{x,L\to\infty \\ x/(2L)=r}} \avg{\Delta_2(\infty)}_a^2 = \begin{cases}
        0& r<1/2\\
        {1-|\!\spr{\Psi}{\Psi'}|^2} & r>1/2
    \end{cases}.
\end{equation}
\begin{figure}\hspace{-1cm}
    \centering
    \includegraphics[width=0.9\linewidth]{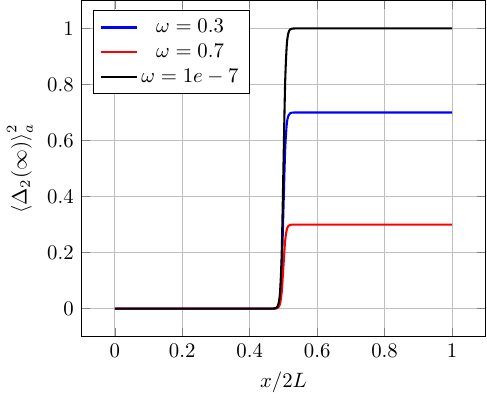}
    \caption{Infinite time limit of $\avg{\Delta_2(t)^2}_a$ given by Eq.~\eqref{eq:long_time_full} for the W-states in Eq.~\eqref{eq:W_state_2} as a function of $x$ with $2L=200$ and for different values of $\omega$ (cf.~Eq.~\eqref{eq:omega}). We see a sharp transition from a regime where memory of the initial state is lost $x/L<1$  to where it is retained for $x/L>1$.}
    \label{fig:inftimegraph}
\end{figure}
An example of this behaviour is shown in Fig.~\ref{fig:inftimegraph}. This confirms our expectations: if the size of the subsystem is smaller than half of the system it eventually looses all the memory of the initial state, however, if it is larger than half of the system the memory of the initial state is never lost. It is worth pointing out that the overlap in the $r>1/2$ regime is taken in the full state, not just the subsystem. This means that for $L\to \infty$ the reduced states can become either identical or maximally distinct (cf.\ Eq.~\eqref{eq:boundsDelta}). We comment further on this below.

\subsubsection{Mixed states}

Up to now we have assumed that our initial state is pure, however, since the main thrust of our calculation amounts to determining the downward evolution of the top boundary condition we can easily accommodate mixed initial states. Moreover, as we work in the folded replicated basis in which density matrices, pure or mixed are represented as vectors the only modification will be that the contraction of the initial state with the all $\squareSA$ state is non trivial. In particular, denoting the initial purity of the states by
\begin{equation}
  \tr[\rho(0)^2]=q^{-2L s},\qquad{\rm tr}[{\rho'}(0)^2]=q^{-2L {s'}}~,
\end{equation}
where $0 \leq s,s'\leq 1$, we have
\begin{equation}
 \label{eq:long_time_mixed}
  \mkern-16mu
  \lim_{t\to\infty}\!\!\avg{\Delta_2(t)}_a^2 \simeq 1\!-\!
  \frac{2 q^{-x}+2 \tr[\rho(0)\rho'(0)] q^{-2L+x}}
  {2 q^{-x}+(q^{-2Ls}+q^{-2Ls'})q^{-2L+x}}.
  \mkern-16mu
\end{equation}
This form is analogous to the pure state result in Eq.~\eqref{eq:long_time_full}, the only difference being the presence of the initial state purities in the denominator. This has the effect of shifting where the initial state memory loss occurs. Using the fact that 
\be 
2\tr[\rho(0)\rho'(0)]\leq \tr[\rho^2(0)]+{\rm \tr}[{\rho'}^2(0)],
\ee 
we find that memory of the initial state can only be retained for 
\begin{equation}
    r>\frac{1+\min[{s,s'}]}{2}. 
\end{equation}
Thus, being mixed makes it more difficult for a system to retain memory of its initial state with larger subsystems required to achieve this. This can be understood intuitively by imagining to purify the mixed states: this effectively increases the size of $\bar A$ keeping $A$ fixed. 

\subsection{Finite-time dynamics}
\label{sec:IntermediateTime}

Let us now move on to illustrate the time evolution of $\avg{\Delta_2(t)}_a$. To this end, we consider two representative examples distinguished by the scaling of the overlap between $\ket{\Psi}$ and $\ket{\Psi'}$ with the volume. 

\subsubsection{Exponential scaling}
We begin considering the example studied in Ref.~\cite{klobas2025translation} of pair product states with four site translational symmetry,
\begin{equation}
 \ket{\Psi} = \bigotimes_{x=1}^{L}\ket{m_x}, \quad \ket{m_x} = \sum_{i,j=1}^q[m_x]_{ij}\ket{i}_x\otimes\ket{j}_{x+\frac{1}{2}}, 
 \label{eq:pairproductstate}   
\end{equation}
where $\ket{m_x}=\ket{m_o}$ for $x$ odd and $\ket{m_e}$ for $x$ even with $m_{e,o}$ being two $q\times q$ unitary matrices and $\ket{i}_x$ is a basis element of the qudit at site $x$. We take $\ket{\Psi'}$ to be related to $\ket{\Psi}$ by a periodic shift of two sites, i.e., it is written as in Eq.~\eqref{eq:pairproductstate} but with $m_e$ and $m_o$ swapped. Note here that these states are mapped into one-another by a two-site shift, which is a symmetry (on average) of our random unitary circuit, so our distance going to zero is a valid indicator of the restoration of this symmetry. In our folded graphical representation we depict the state at each pair of sites as
\begin{equation}
\ket{M_x}=\ket{m_x}\otimes\ket{m_x}^*\otimes\ket{m_{x-1}}\otimes\ket{m_{x-1}}^*=
    \begin{tikzpicture}[baseline={(0.2,-.25)},scale=0.75]
  \istate{0}{0}{colSt},
  \end{tikzpicture}
\end{equation}
and therefore write
\begin{equation}
    \tr[\rho_a(t)\rho_A'(t)] = 
\begin{tikzpicture}[baseline={(0,-1.5)},scale=0.5]
    \foreach \x in {-9,-7,...,-1}{\prop{\x}{0}{favg}{}}
   \foreach \x in {-8,-6,...,-2}{\prop{\x}{-1}{favg}{}}
   \foreach \x in {-9,-7,...,-1}{\prop{\x}{-2}{favg}{}}
   \foreach \x in {-8,-6,...,-2}{\prop{\x}{-3}{favg}{}}
   \foreach \x in {-9,-7,...,-1}{\prop{\x}{-4}{favg}{}}
   \foreach \x in {-8,-6,...,-2}{\prop{\x}{-5}{favg}{}}
   \foreach \x in {-9.5,...,-5.5}{\circle{\x}{0.5}}
   \foreach \x in {-4.5,...,-0.5}{\square{\x}{0.5}}
   \foreach \x in {-9,-7,...,-1}{
   \istate{\x-.5}{-5.5}{colSt}
   }
   
   \draw[color = flines, line width = 0.2mm](-9.5,-.5) -- (-9.5,-1.5);
   \draw[color = flines, line width = 0.2mm](-.5,-2.5) -- (-.5,-3.5);
   \draw[color = flines, line width = 0.2mm](-9.5,-2.5) -- (-9.5,-3.5);
   \draw[color = flines, line width = 0.2mm] (-.5,-.5) -- (-.5,-1.5);
   \draw[color = flines, line width = 0.2mm] (-9.5,-4.5) -- (-9.5,-5.5);
   \draw[color = flines, line width = 0.2mm] (-.5,-4.5) -- (-.5,-5.5);
    
    \end{tikzpicture}
\end{equation}
To compute this diagram, there are four quantities we must evaluate.
\begin{equation}
  \mkern-16mu
  \begin{aligned}
    \spr{\circleSA\,\circleSA}{M_x} &= 
    \begin{tikzpicture}[baseline={(0.2,-0.25)},scale=0.65]
      \istate{0}{0}{colSt}
      \circle{0}{0}
      \circle{1}{0}
    \end{tikzpicture} = 1, &
    \spr{\squareSA\,\squareSA}{M_x} &= 
    \begin{tikzpicture}[baseline={(0.2,-0.25)},scale=0.65]
      \istate{0}{0}{colSt}
      \square{0}{0}
      \square{1}{0}
    \end{tikzpicture} = \beta^2,
    \\
    \spr{\circleSA\,\squareSA}{M_x} &= 
    \begin{tikzpicture}[baseline={(0.2,-0.25)},scale=0.65]
      \istate{0}{0}{colSt}
      \circle{0}{0}
      \square{1}{0}
    \end{tikzpicture}=\gamma,&
    \spr{\squareSA\,\circleSA}{M_x} &= 
    \begin{tikzpicture}[baseline={(0.2,-0.25)},scale=0.65]
      \istate{0}{0}{colSt}
      \circle{1}{0}
      \square{0}{0}
    \end{tikzpicture} = \gamma,
  \end{aligned}
  \mkern-16mu
\end{equation}
where $\beta,\gamma\in[0,1]$ are parameters set by our choices of $m_e$ and $m_o$, and are given by
\begin{equation}
  \beta=|\mathrm{tr}(m_em_0^\dagger)|,\qquad
  \gamma=\tr(m_em_e^\dagger m_om_o^\dagger).
\end{equation}
This immediately implies
\begin{equation}
  \braket*{\circleSA^{x}\,\squareSA^{2L-x}}{\Psi_4\brakapp} 
  = \beta^{x}\left(\frac{\gamma}{\beta}\right)^p
\end{equation}
where $p=x\pmod{2}$. For simplicity, we can take $\gamma=\beta$ to remove this dependence on parity without changing the underlying physics. Note also that in this situation we have $\spr{\Psi_4}{\Psi_4\brakapp}=\beta^{2L}$. Using these facts, we obtain the exact time dependence. To simplify the notation we introduce $\O_C(m,\beta),\O_S(m,\beta),\O_{X}(m,\beta),$ which denote the contributions to $\braket{T_0(x)}{\smash{\Psi_4\brakapp}}$ from stationary, circle ($C$) and  square ($S$) states and the non-stationary states $(X)$.  These are given by 
\begin{equation}
  \mkern-16mu
\begin{aligned}
  \mathcal{O}_C(m,\beta)&=A_m(x),\qquad 
  \mathcal{O}_S(m,\beta)=\beta^{2L}A_m(2L-x),\\ 
  \mathcal{O}_X(m,\beta)&= \alpha^m\beta^{x-m}
  \smashoperator{\sum_{k=k_{\min}}^{k_{\max}}} C_{k,m}(x)\beta^{2k},
\end{aligned}
  \mkern-16mu
\end{equation}
and with this the final result is
\begin{equation}
\label{eq:annealeddisteq}
   \avg{\Delta_2(t)}_a^2 = 
   1-\frac{\O_X(2t,\beta)+\O_C(2t,\beta)+\O_S(2t,\beta)}{\O_X(2t,1)+\O_C(2t,1)+\O_S(2t,1)}.
\end{equation}

\begin{figure}\hspace{-1cm}
    \centering
    \includegraphics[width=0.9\linewidth]{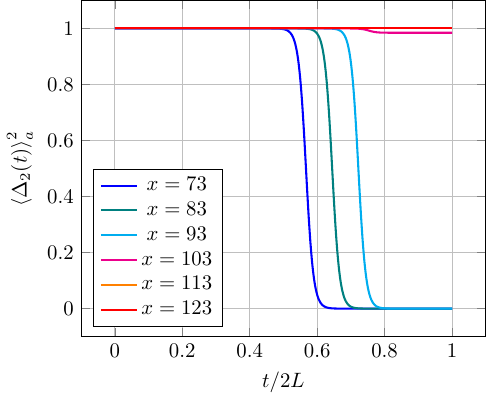}
    \caption{Time dependence of $\avg{\Delta_2(t)^2}_a$ from the initial pair product state in Eq.~\eqref{eq:pairproductstate} and different subsystem sizes $x$. We take $2L=200,q=2,\beta=0.7$. For $x<L$ the curves drop to zero while for $x>L$ they stay finite for all times. The difference between the curves for the two largest values of $x$ is not distinguishable on this scale.  }
    \label{fig:timedependencepairproduct}
\end{figure}

We plot this function for different subsystem sizes in Fig.~\ref{fig:timedependencepairproduct} and see the expected long time behaviour given by Eq.~\eqref{eq:long_time_behaviour}. To understand the intermediate time behaviour, we note that  when ${2t<\min(x,2L-x)}$ the dynamics is not affected by the boundary conditions. For our purposes, this has two main consequences. Firstly, it means there is no contribution from the \emph{all} $\circleSA$ or \emph{all} $\squareSA$ states, and so we are left only with decaying contributions. Secondly, since our boundary conditions can now be ignored, our recursion relation simply yields a binomial coefficient $C_{k,m}(x) = {m\choose k}$. Using these we can generate a simple analytic expression for our distance for short time. Specifically, noting that
\begin{equation}
  \begin{aligned}
    \O_X(2t,\beta) &= \alpha^{2t}\sum_{k=0}^{2t}{2t\choose k}\beta^{x-2t+2k}\\ 
    &= \alpha^{2t}\beta^{x-2t}(1+\beta^2)^{2t}
  \end{aligned}
\end{equation}
we obtain 
\begin{equation}
   \avg{\Delta_2(t)}_a^2 = 1-\beta^{x}\left(\frac{1+\beta^2}{2\beta}\right)^{2t}\mkern-12mu ,
   \quad
   t<\min(x,2L-x), 
\end{equation}
which, as expected, is in agreement with the short time behaviour of the infinite-$L$ limit~\cite{klobas2025translation}. When $x<L$ but $2t>x$ we see from the figure that $\langle\Delta_2(t)^2\rangle_a$ rapidly drops to zero. 

Note that at time zero we have $\Delta_2(0)^2=\avg{\Delta_2(0)}_a^2=1-\beta^x$, which is less than our result at infinite time for $x>L$. This means that in this case the distance actually increases, albeit that this is exponentially suppressed. 

Lastly, we stress that this argument can be naturally extended for any set of initial \emph{$s$-site product} states $\ket{\Psi}$, $\ket{\Psi'}$, with finite $s$, %support size $s$,
since we will have the overlap of such states $|\spr{\Psi}{\Psi'}|^2$ exponentially suppressed with the system size, and as such the behaviour we see with our two-site product state is qualitatively similar for all short range entangled states. 

\subsubsection{Constant overlap}

Let us now consider the example of a state with long-range entanglement. Specifically, let us look at two-site shift invariant $W$ states defined as
\begin{equation}\label{eq:W_state}
    \ket{w} = \frac{1}{\sqrt{L}}\sum_{n=0}^{L-1}(c_1+c_2(-1)^n)\ket{d_n},
\end{equation}
where 
\be
\ket{d_n} = |\overbrace{0\cdots0}^{2n}10\cdots0\rangle, 
\ee
and $c_{1,2}$ are complex numbers satisfying the normalization $|c_1|^2+|c_2|^2=1$ and the local Hilbert space dimension is $q=2$. These states are generalizations of the standard $n$-qubit W states, which are one-site translationally invariant. Alternatively, they can be viewed as single magnon states in the language of spin chains. They possess $\log(2)$ half system entanglement entropy and have also been studied in the context of multipartite entanglement~\cite{dur2000three,park2025entangled}.

Taking $\ket{\Psi}=\ket{w}$ and $\ket*{\Psi^{\prime}}=\ket*{w^{\prime}}$ its two-site shifted counterpart,
\begin{equation} \label{eq:W_state_2}
  \ket{w'} = \frac{1}{\sqrt{L}}\sum_{n=0}^{L-1}(c_1-c_2(-1)^n)\ket{d_n},
\end{equation}
we find that the overlap between $\ket{\Psi}$ and $\ket{\Psi'}$ is in fact independent of system size and is given by
\be\label{eq:omega}
\omega\equiv|\spr{w}{w'}|^2 = (|c_1|^2-|c_2|^2)^2\,. 
\ee
This means that in the thermodynamic limit and for $x>L$ the states do not simply become maximally different: some non-trivial initial-state dependence remains. Specifically, in the infinite-time limit the distance goes to
\be
\avg{\Delta_2(\infty)}^2 =1-\omega,
\ee
which means that we maintain a finite amount of information, $\omega$, about our original states. 

\begin{figure}\hspace{-1cm}
    \centering
    \includegraphics[width=0.9\linewidth]{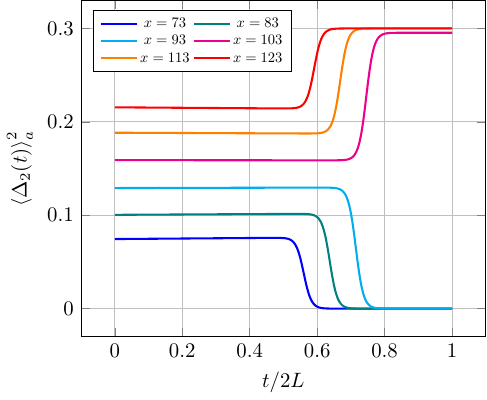}
    \caption{Time dependence of $\avg{\Delta_2(t)^2}_a$ from the pair of states in~\eqref{eq:W_state_2} for different subsystem sizes $x$. We take $2L=200,q=2,\omega=0.7$.}
    \label{fig:timedependencewpair}
\end{figure}

To understand the dynamics we first look at short times $2t<\min(x,2L-x)$. In this regime one can again calculate the distance by replacing the coefficients $C_{k,m}(x)$ by binomials, see App.~\ref{sec:shorttime}. The result reads 
\begin{equation}
\label{eq:wdistshorttime}
    \avg{\Delta_2(t)}_a^2 = 1-\frac{(2L-x)^2+\omega x^2 + 2(1+\omega)t}{(2L-x)^2+x^2+4t},
\end{equation}
and now features finite-time deviations that are suppressed algebraically, rather than exponentially as before. Also note that 
\be
\frac{(2L-x)^2+\omega x^2}{(2L-x)^2+x^2} \geq \frac{\omega (2L-x)^2+\omega x^2}{(2L-x)^2+x^2}=\omega,
\ee 
which means that $\avg{\Delta_2(0)}_a^2\leq  \avg{\Delta_2(\infty)}_a^2$ for \(x>L\). We plot the full time dynamics for different values of $x$ in Fig.~\ref{fig:timedependencewpair}.

Now we can combine Eq.~\eqref{eq:wdistshorttime} with the $t\to\infty$ result in Eq.~\eqref{eq:long_time_behaviour} to obtain an asymptotic expression for the annealed average distance in the thermodynamic limit. Considering for instance $x<L,\; \omega=0$, we have 
\be
\tr\left[\rho_A(t)\rho'_A(t)\right]\simeq \frac{\alpha^{2t}}{(2L)^2}((2L-x)^2+2t)+q^{-x}, 
\ee
and
\be
\tr\left[\rho_A(t)\rho_A(t)\right] \simeq \frac{\alpha^{2t}}{(2L)^2}((2L-x)^2+x^2+4t)+q^{-x}.
\ee
We see that for time scales where the short time result is relevant ($t=O(L)$), this linear time term contributes as $O(1/L)$, and so in the large system limit we can simply drop it at the cost of algebraically small errors. With this and some rearranging one gets an annealed average distance of 
\be
\!\!\!\avg{\Delta_2(t}_{a}^2|_{\omega=0}   \simeq \frac{r^2}{(1-r)^2+r^2+ q^{-(x-v_et)}},
\ee
where $r=x/2L$ and $v_e = -2\ln(2\alpha)/\ln(q)$ is the standard entanglement velocity. One can implement a similar procedure for the case where $x>L$. For a generic $\omega$, it is fairly simple to show this just has the effect of scaling our distance by a factor of $1-\omega$. So, overall, we find the following asymptotic expression
\begin{equation}
 \!\!\! \!\!\! \frac{\avg{\Delta_2(t)}_a^2}{(1-\omega)} \simeq
  \begin{cases}
   \frac{r^2}{({1}-r)^2+r^2+ q^{-(x-v_et)}}, &  x<L,\\
    1-\frac{(1-r)^2}{(1-r)^2+r^2+q^{x+v_et-2L)}}, &  x>L.
  \end{cases}
  \label{eq:asyDelta}
\end{equation}

\section{Open systems}\label{sec:OpenDynamics}
Let us now consider the situation where there is dissipation at the edge of the circuit, modelling the presence of noise or of a coupling to an external environment. In contrast to the unitary dynamics of the bulk, the dissipation will force the system to lose memory of the initial state. For finite dissipation in a finite system all memory of the initial state will eventually be lost regardless of subsystem size. The expectation is, however, that, for large enough subsystems, tuning the dissipation in time one can explore a phase transition between a memory retaining phase and one where the information is completely lost. This kind of transition, dubbed ``quantum coding transition'', has recently been studied in Ref.~\cite{lovas2024quantum}.  

To proceed we introduce a boundary dissipation operator defined as 
\begin{equation}
  D = (1-p)\cdot \mathbbm{1} +\frac{p}{q^2}\ket{\circleSA}\bra{\circleSA} = 
  \begin{tikzpicture}[baseline={(0,.4)},scale=0.7]
    \draw[color=flines](0,0) -- (0,1);
    \obs{0}{.5}{myred}
  \end{tikzpicture},
\end{equation}
for some parameter $p\in[0,1]$. The action of $D$ on $\ket{\circleSA},\ket{\squareSA}$ is thus given by 
\begin{equation}
\begin{aligned}
&D\ket{\circleSA} = 
    \begin{tikzpicture}[baseline={(0,.25)},scale=.7]
        \draw[color=flines](0,0) -- (0,1);
        \obs{0}{.5}{myred}
        \circle{0}{1}
    \end{tikzpicture} = \ket{\circleSA},\\
 &D\ket{\squareSA} = 
    \begin{tikzpicture}[baseline={(0,.25)},scale=.7]
        \draw[color=flines](0,0) -- (0,1);
        \obs{0}{.5}{myred}
        \square{0}{1}
    \end{tikzpicture} = (1-p)\ket{\squareSA}+\frac{p}{q}\ket{\circleSA}.
\end{aligned}
\end{equation}
We apply this dissipation operator instead of the identity along the edge away from our subsystem $A$. In this way our quantity of interest becomes 
\begin{equation}
    \tr[\rho_A(t)\rho_A\brakapp(t)] = 
    \begin{tikzpicture}[baseline={(0,-1.5)},scale=0.5]
    \foreach \x in {-9,-7,...,-1}{\prop{\x}{0}{favg}{}}
   \foreach \x in {-8,-6,...,-2}{\prop{\x}{-1}{favg}{}}
   \foreach \x in {-9,-7,...,-1}{\prop{\x}{-2}{favg}{}}
   \foreach \x in {-8,-6,...,-2}{\prop{\x}{-3}{favg}{}}
   \foreach \x in {-9,-7,...,-1}{\prop{\x}{-4}{favg}{}}
   \foreach \x in {-8,-6,...,-2}{\prop{\x}{-5}{favg}{}}
   \foreach \x in {-9.5,...,-5.5}{\circle{\x}{0.5}}
   \foreach \x in {-4.5,...,-.5}{\square{\x}{0.5}}
   \draw[color = flines, line width = 0.2mm](-9.5,-.5) -- (-9.5,-1.5);
   \draw[color = flines, line width = 0.2mm](-.5,-2.5) -- (-.5,-3.5);
   \draw[color = flines, line width = 0.2mm](-9.5,-2.5) -- (-9.5,-3.5);
   \draw[color = flines, line width = 0.2mm] (-.5,-.5) -- (-.5,-1.5);
   \draw[color = flines, line width = 0.2mm] (-9.5,-4.5) -- (-9.5,-5.5);
   \draw[color = flines, line width = 0.2mm] (-.5,-4.5) -- (-.5,-5.5);
    \obs{-9.5}{-3}{myred}
    \obs{-9.5}{-5}{myred}
    \obs{-9.5}{-1}{myred}
   \draw[color = flines, line width = 0.2mm, fill=colSt] (-9.65,-6.6) rectangle (-.35,-5.5)
   node[pos=.5] {$\Psi_4\brakapp$};
    \draw[semithick,decorate,decoration={brace}] (-4.65,.76) -- (-.35,.76) node[midway,xshift=0pt,yshift=7.5pt] {$A$};
    \end{tikzpicture}.
\end{equation}
%For completeness, in this section we will continue using our inhomogeneous Dicke state, $\ket{w_{ab}}$, so that the increase in asymmetry is not exponentially suppressed, but the results will essentially be general. \bruno{I suggest to move this comment later}
For any fixed rate of dissipation $p=O(1)$ the initial-state memory will eventually be entirely lost even in the cases where our subsystem consists of over half our total system. Therefore, the most interesting case is that of a dissipation that weakens in time. Equivalently, one can consider the evolution for a fixed number of time steps, or circuit depth, $T$, and take $p=O(1/T^\beta)$ for some $\beta>0$. This approach is more convenient because it allows us to characterise the effect of dissipation using perturbation theory in $1/T^\beta$. Before doing this, we proceed by performing a convenient rewriting of the ``downward evolution'' in averaged circuit, in terms of a Markov matrix.

\subsection{Markov Matrix Formalism}

Focussing on the non-dissipative case ($p=0$) consider the expansion Eq.~\eqref{eq:Tstateexpanded} of the state $\bra{T_{m}(x)}$ in terms of ``domain wall'' states 
\be
 \bra{T_0(y)} = \bra{\circleSA^{2L-y}\squareSA^y}, 
\ee
and note that for odd (even) $m$ only domain walls at even (odd) positions appear apart from the two domains of circle and square states.
Focussing on even $m$ --- we are interested in a circuit evolving for an integer number of full time steps (cf.\ Eq.~\eqref{eq:timeevolutionop}) ---  we then have $L+2$ independent states and we can arrange their coefficients in an $(L+2)$-dimensional vector $W_m$ defined as follows: the first entry is the coefficient of $\bra{T_0(0)}$, the $\ell$-th entry, for $\ell=2,\ldots,L+1$, is the coefficient of $\bra{T_0(2\ell-3)}$, and the $(L+2)$-th entry is the coefficient of $\bra{T_0(2L)}$. The evolution equation for $\bra{T_{m}(x)}$ between even times can then be written as
\begin{equation}
\label{eq:markoveq}
    W_{2m+2} =  A_0 W_{2m},
\end{equation}
where $A_0$ is a $(L+2)\times (L+2)$ Markov matrix given by 
\be
A_0= \begin{pmatrix}
& 1 & \alpha & & & \\
 & & \alpha^2 & \alpha^2 & &\\
    & & \alpha^2 & 2\alpha^2& \alpha^2 &\\
    & &  &\ddots & \ddots & \ddots \\
    & & & & \alpha^2  & 2\alpha^2 & \alpha^2 \\
    & & & && \alpha^2 &\alpha^2\\
    & & & && & \alpha & 1 &
\end{pmatrix}
\ee
 The matrix $A_0$ has two eigenvalues $1$, corresponding to the right eigenvectors $R_1 = (1,0,\ldots,0)$ and $R_{L+2} = (0,\ldots,0,1)$ and all its other eigenvalue have magnitude strictly smaller than $1$. The corresponding left eigenvectors are denoted $L_1,L_{L+2}$. Solving Eq.~\eqref{eq:markoveq} as 
\be
W_{2m} = A_0^{m} W_{0},
\ee
we see that the eigenvalue structure of $A_0$ is consistent with what we found in Sec.~\ref{sec:exactannealedaverage}: Only the states $\bra{T_0(0)}$ and $\bra{T_0(2L)}$ survive at infinite times.

Let us now add dissipation. Following our previous discussion we consider  
\be
p(a) = a\left(\frac{L}{T}\right)^\beta,
\ee 
where $a,\beta>0$ and, once again, $T$ is the depth (number of evolution steps) of our circuit. This modifies the Markov matrix in Eq.~\eqref{eq:markoveq} to
\be
A_0 \mapsto A = A_0 + a\left(\frac{L}{T}\right)^\beta P,
\ee
where
\begin{equation}
    P = \begin{pmatrix}
        &\vdots&\vdots&\vdots\\
        \cdots&0&0&0\\
        \cdots&0&\alpha/q&1/q\\
        \cdots&0&-\alpha&-1

    \end{pmatrix}\,.
\end{equation}
The vector $R_1$ continues to be an eigenvector of $A$ with eigenvalue one, however, the other eigenvalue one is reduced by the perturbation. Specifically, as we show in Appendix~\ref{sec:perturbation}, we have 
\be\label{eq:shiftedeigenvalue}
\lambda=1-a\left(\frac{L}{T}\right)^\beta\left(\frac{1-q^{-2}}{1-q^{-2L}}\right) + o\left(\frac{1}{T^\beta}\right)
\ee 
This means that, choosing $\beta=1$, we have 
\be
\lim_{T\to\infty} \!\!\!W_{T} \simeq R_1 (L_1\!\cdot\! W_0) \!+\! R_{L+2} (L_{L+2}\!\cdot\! W_0) e^{-{a}L\left(\frac{q^2-1}{q^2}\right)}. 
\ee
Thus, the time evolution operator becomes a projector onto the all $\circleSA$ or all $\squareSA$ states with the latter being exponentially suppressed. Depending on the initial state and the subsystem size, however, this suppression can be overcome. To see this in action, let us take as our initial states the two site translationally invariant W-states~\eqref{eq:W_state_2}. With these, we find 
\be
\label{eq:open_long_time}
\avg{\Delta_2(T)}^2 = 1-\frac{q^{-x}+\omega e^{-aL(1-1/q^2)}q^{-2L+x}}{q^{-x}+e^{-aL(1-1/q^2)}q^{-2L+x}}, 
\ee
which is reminiscent of the expression for the long time behaviour of mixed states reported in Eq.~\eqref{eq:long_time_mixed}. That is, the second term in the denominator has gained an additional exponential dependence. Accordingly, we see that, even for $x>L$ there are values of $a$ for which $\avg{\Delta_2(T)}^2=0$, in line with our intuition. This is not always the case, however, and there 
exists an $a_c$ such that if $ a< a_c$ memory of the initial state is retained. For a fixed $x$ this value is given by 
\begin{equation}
 a_c = \frac{(2r-1) q^2\ln q}{q^2-1}, 
\end{equation}
where we recall $r = x/{(2L)}$ denotes the proportion of our system constituted by the subsystem (cf.\ Eq.~\eqref{eq:asyDelta}). Conversely, one can instead find 
\be
r_c = \frac{1}{2}+\frac{a(q^2-1)}{2q^2\ln q},
\ee
to get the minimum size our subsystem can be and still retain memory of the initial state for a fixed amount of noise.  Note that in the limit of large Hilbert space dimension $r_c\to 1/2$, returning to the unitary value.  
\begin{figure}[t]
    \centering\hspace{-1.2cm}
    \includegraphics[width=0.9\linewidth]{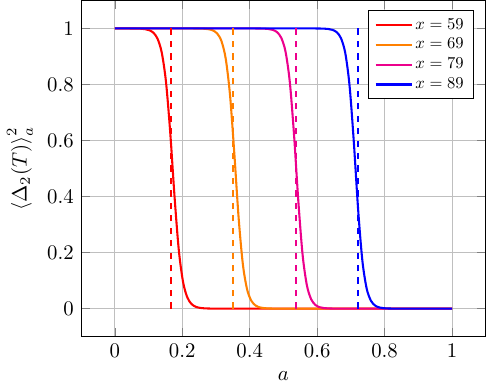}
    \caption{Values of $\avg{\Delta_2(T)}^2$ in the presence of boundary dissipation for the W states from \eqref{eq:open_long_time} as a function of dissipation strength $a$  and for different values of subsystem size. We take $q=2,2L=100,T=20L$ and $\omega=0.7$. Dashed lines indicate value of $\Bar{a}_c$ predicted with first order perturbation theory.}
    \label{fig:placeholder}
\end{figure}

\section{Conclusions}\label{sec:conclusions}

In this paper we have studied the behaviour of initial-state memory under local random unitary dynamics. In particular, we have compared the reduced density matrices obtained by evolving two different initial states with the same realisation of a random brickwork circuit, averaging their distance over different realisations. 

We have shown that if the subsystem is smaller than half of the full system the two reduced states become exponentially close in a time-scale that grows linearly with the subsystem size following a universal curve. On the other hand, if the subsystem is larger than half of the full system, the memory of the initial state is never lost, and for sufficiently entangled pairs of initial states the distance gets amplified in a time-scale that grows linearly with the size of the complement. Once again, the amplification dynamics becomes universal at large scales. Interestingly, if the initial state is mixed, the subsystem size at which the transition between memory loss and retaining happens is larger than half of the system size. Finally, we have also considered the effect of the (boundary) dissipation on the memory loss. We have shown that if one adds a finite amount of dissipation for each time-step, the memory of the initial state is eventually lost regardless of the subsystem size. However, if the strength of the dissipation is appropriately reduced in time, one can find a non-trivial transition between memory loss and retention.

These results show that, even though chaotic quantum circuits are expected to rapidly scramble local information, the reduced states of subsystems retain information of the initial state up to long times (in fact, up to infinite times for large enough subsystems) even in the extreme case of random unitary circuits. It would be interesting to understand how this phenomenology, and the universal restoration dynamics that we observed, changes when the dynamics becomes more constrained and the memory effects are expected to be stronger, e.g., when they feature non-trivial conservation laws~\cite{liu2024symmetry, turkeshi2025quantum}. 

\begin{acknowledgments}
We acknowledge financial support from the Royal Society through the University Research Fellowship No.\ 201101 (B.\ B.\ and J.\ B.).  
\end{acknowledgments}

\appendix 

\section{Simplification of $A_{\infty}(x)$}
\label{sec:AppendixAinfty}

The coefficient $A_{\infty}(x)$ is given by 
\begin{equation}
  \mkern-16mu
  \begin{aligned}
    A_\infty(x) = \alpha^x\sum_{k=0}^\infty\alpha^{2k}
    &\left[{2k+x-1\choose k}-{2k+x-1\choose k+x}\right.\\
    &\left. {\vphantom{{2k+x-1\choose k}}} -r_{k,2k+x-1}(x) \right].
  \end{aligned}
  \mkern-16mu
\end{equation}
We will simplify this expression by taking into account the following identity,
\be
\sum_{n=0}^\infty{2n+a\choose n}z^n = \frac{G(z)^a}{\sqrt{1-4z}},
\ee
where we set 
\be
G(z) = \frac{1-\sqrt{1-4z^2}}{2z}.
\ee
Noting that 
\begin{equation}
  G(\alpha^2)=\alpha^{-1}q^{-1},\qquad\frac{1}{\sqrt{1-4\alpha^2}}=\frac{q^2+1}{q^2-1},
\end{equation}
we can evaluate the first term as 
\begin{equation}
  \mkern-16mu
    \alpha^x\sum_{k=0}^\infty\alpha^{2k}{2k+x-1\choose k} 
    = \alpha^x\frac{G(\alpha^2)^{x-1}}{\sqrt{1-4\alpha^2}} = \frac{q^{2-x}}{q^2-1}.
  \mkern-16mu
\end{equation}
The second term can be calculated in the same fashion
\begin{equation}
  \mkern-16mu
\begin{aligned}
  &\alpha^x\sum_{\rho=0}^{\infty}\alpha^{2\rho}{2\rho+x-1\choose\rho+x} 
  = \alpha^x\sum_{\rho=1}^{\infty}\alpha^{2\rho}{2\rho+x-1\choose\rho-1}\\ 
  &=\alpha^{x+2}\sum_{r=0}^{\infty}\alpha^{2r}{2r+x+1\choose r} = \frac{q^{-x}}{q^2-1}.
\end{aligned}
  \mkern-16mu
\end{equation}
Thus we find
\begin{equation}
A_\infty(x) =q^{-x}-\alpha^x\sum_{\rho=0}^\infty\alpha^{2\rho}r_{\rho,2\rho+x-1}(x).
\end{equation}
The remaining sum can be computed by recalling the definition of $r_{k,m}(x)$ (cf.\ Eq.~\eqref{eq:rdef}). This allows us to rewrite the sum as  
\begin{equation}
\!\!\!\!\!\begin{aligned}
\alpha^x\sum_{n=1}^\infty\sum_{\rho=0}^{\infty}\alpha^{2\rho}
  &\left[{2\rho+x-1\choose \rho+x+2nL}+{2\rho+x-1\choose \rho+x-2nL}\right.\\
  &\left.-{{2\rho+x-1\choose \rho+2nL}-{2\rho+x-1\choose \rho-2nL}}\right].
\end{aligned}
\end{equation}
Using a procedure analogous to the one presented above, we can also determine
\be
\!\!\!\!\alpha^x\sum_{\rho=0}^\infty\alpha^{2\rho}{2\rho+x-1\choose \rho + \nu x \pm 2nL} = q^{\pm(-1)^\nu x-4nL}\frac{q^{1\mp1}}{q^2-1},
\ee
where \(\nu=0,1\). Therefore, we find that the subleading term is equal to
\begin{equation}
  \begin{aligned}
    \alpha^x\sum_{\rho=0}^\infty\alpha^{2\rho}r_{\rho,2\rho+x-1}(x)
    &= \sum_{n=1}^\infty q^{-4nL}(q^x-q^{-x})\\
    &=\frac{2\sinh{(x\ln q)}}{q^{4L}-1}.
  \end{aligned}
\end{equation}
Collecting the different pieces, we finally arrive at the expression
\begin{equation} \label{eq:AinfApp}
  A_\infty(x) = q^{-x} -\frac{2\sinh{(x\ln q)}}{q^{4L}-1}
\end{equation}

\section{Downward evolving state for an all-to-all random unitary matrix}
\label{sec:alltoall}

It is known that for legs of arbitrary dimension $p,q$, that our local gate acts as
\be
\begin{aligned}
    \aU_2^{pq}\ket{\circleSA_p\squareSA_q} &= \frac{1}{q^2p^2-1}(Q_p\ket{\circleSA_p\circleSA_q}+Q_p\ket{\squareSA_p\squareSA_q})\\ &= A_\infty\ket{\circleSA^{2L}} + B_\infty\ket{\squareSA^{2L}}
\end{aligned}
\ee
where $Q_p=(p^2-1)q$ and $Q_q=(q^2-1)p$. Setting $p\mapsto q^{2L-x}$, $q\mapsto q^x$ one can directly determine,
\begin{equation}
\begin{aligned}
    &A_\infty = q^x\frac{q^{2(2L-x)}-1}{q^{4L}-1} = q^{-x}-\frac{2\sinh(x\ln q)}{q^{4L}-1}\\
    &B_\infty = q^{-2L+x}-\frac{2\sinh((2L-x)\ln q)}{q^{4L}-1},
\end{aligned}
\end{equation}
in agreement with Eqs.~\eqref{eq:Ainf} and \eqref{eq:Binf}.

\section{Short-time behaviour of \(\ket{w}\) states.}
\label{sec:shorttime}
To calculate the time evolution of \(\avg{\Delta_2(t)}_a^2\) for our \(\ket{w}\) states, we must evaluate terms of the form \(\spr{\circleSA^{2L-x}\squareSA^x}{\Psi_4'}\), where \(\Psi_4'=\ket{w}\otimes\ket{w}^*\otimes\ket{w'}\otimes\ket{w'}^*\) for an arbitrary \(x\). Writing this out explicitly we get that this overlap is equal to
\be
\frac{1}{L^2}\sum_{n_{1,2,3,4}=0}^{L-1} R_{1234} \spr{\circleSA^{2L-x}\squareSA^x}{d_{n_1}d_{n_2}d_{n_3}d_{n_4}}
\ee
where
\begin{equation}
\begin{aligned}
R_{1234} = (c_1+&c_2(-1)^{n_1})(c_1^*+c_2^*(-1)^{n_2})\\ &(c_1-c_2(-1)^{n_3})(c_1^*-c_2^*(-1)^{n_4}).
\end{aligned}
\end{equation}
To evaluate this sum we determine what values of \(n_{1,2,3,4}\) give a non-zero contribution. For instance if \(n_1<\floor{L-x/2}\), then our \(\ket{d_{n_1}}\) term is contracted with \(\ket{d_{n_2}}\) via a bullet state, and so we must have \(n_1=n_2\) to get a non-zero entry. Continuing for all cases we find that the only non-zero terms come from when either \(n_1,n_3\) both coincide with bullet states and \(n_1=n_2,n_3=n_4\), or both coincide with square states and \(n_1=n_4,n_3=n_2\). Thus our total overlap can be decomposed as 
\be
L^2\spr{\circleSA^{2L-x}\squareSA^x}{\Psi_4'} = \sum_{n_{1,3}=1}^{\floor{L-x/2}} R_{1133} + \sum_{n_{1,3}=1}^{\ceil{x/2}} R_{1331}.
\ee
Focusing for now just on our first term, one can compute that \(R_{1133} = 1+k((-1)^{n_1}-(-1)^{n_3})-k^2(-1)^{n_1+n_3}\) where \(k=2\Re(c_1c_2^*)\), from which we get that this term simply yields \((L-x/2)^2\) plus some potential algebraically suppressed parity terms which become negligible for large systems. Similary one can compute the second sum to be equal to \(\omega(x/2)^2\), again plus some negligible parity terms. Hence overall, we can determine
\be\label{eq:wstateoverlaps}
\spr{\circleSA^{2L-x}\squareSA^x}{\Psi_4'} = \frac{(2L-x)^2+\omega x^2}{(2L)^2}.
\ee
Plugging this into the expression for \(\O_X\) at short time one determines 
\be\label{eq:nonstationarywoverlaps}
\!\!\!\! \O_X(2t,\omega) = \frac{\alpha^{2t}}{(2L)^2}\left((2L-x)^2+\omega x^2+2(1+\omega)t\right).
\ee
With this, and using Eq.~\eqref{eq:annealeddisteq} at short times where stationary overlaps equal zero, one recovers Eq.~\eqref{eq:wdistshorttime}.

\section{Perturbation theory for $A$}
\label{sec:perturbation}

For large $T$ we can then find the eigenvalues of $A$ using perturbation theory, with the only caveat that our eigenvalue $\lambda=1$ is degenerate, and our evolution operator $A_0$ is non-symmetric and so we must consider a bi-orthogonal basis \(\{L_i,R_j\}\) with the condition \(L_iR_j\propto \delta_{ij}\). This lets us write a first-order correction to our eigenvalue associated with $R_{L+2} = (0,\dots,0,1)^\mathsf{T}$, $\lambda_1=1$ as,
\begin{equation}
    \lambda_1^{(1)} = \frac{L_{L+2}PR_{L+2}}{L_{L+2}R_{L+2}}.
\end{equation}
Note since our problem is degenerate we must be careful to select \(L_{L+2}\) orthogonal to \(R_{1}=(1,0,\dots,0)^{\mathsf{T}}\), meaning we must have its first element equal to zero. One can show the desired left eigenvector is equal to
\be\label{eq:lefteigenvector}
L_{L+2} = \left(0,Q_0,Q_2,\dots,Q_{L-4},Q_{L-2},Q_{L-1}\right),
\ee
where
\be
Q_N = \frac{q^{-N}(q^{2N+2}-1)}{q^2-1}.
\ee
This is calculated by noting $L_{L+2}A_0=L_{L+2}$ and so letting \(L_{L+2} = (C_1,\dots,C_{L+2})\) we can set \(C_1=0,C_2=1\), since we require it to be orthogonal to \(R_1\) and have one extra degree of freedom by not requiring \(L_{L+2}\) to be normalised, yielding the system of equations
\be
\mkern-20mu
\begin{aligned}
  C_{L+1}&=\alpha C_{L+2}+\alpha^2(C_{L+1}+C_L), & 1 &= \alpha^2(C_{3}+1),\\
  C_j &= \alpha^2(C_{j-1}+2C_j+C_{j+1}),& 4&\leq j\leq L.
\end{aligned}
\mkern-12mu
\ee
$C_3$ is simply obtained algebraically and can be shown to  $\frac{q^{-2}(q^6-1)}{q^2-1} = Q_2$.
Next observe that $q^2(Q_{N-2}+2Q_N+Q_{N+2}) = Q_N(q^2+1)^2$, and thus our main set of equations is satisfied. For our first equation, note that one can show
\begin{equation}
\alpha Q_{L-1} +\alpha^2Q_{L-4} = (q^4+q^2+1) Q_{L-2},
\end{equation}
and the fact our first equation is true holds immediately. Hence, Eq.~\eqref{eq:lefteigenvector} is the correct left eigenvector.
\par
And so we can get our first order correction
\begin{equation}
\begin{aligned}
 \frac{L_{L+2}PR_{L+2}}{L_{L+2}R_{L+2}} &= -\left(1-\frac{1}{q}\frac{Q_{L-2}}{Q_{L-1}}\right) = -\left(\frac{1-q^{-2}}{1-q^{-2L}}\right),
 \end{aligned}
\end{equation}
using the fact that \(Q_{L-1}/Q_{L-2} = q+\frac{q^2-1}{q^{2L-1}}\). Adding the scaling factor for our perturbation gives the shifted eigenvalue in Eq.~\eqref{eq:shiftedeigenvalue}. 

\bibliography{Restoration.bib}

\end{document}